# Structural properties and Raman spectroscopy of lipid Langmuir monolayers at the air–water interface

Shuxi Dai [a, b], Xingtang Zhang [a], Zuliang Du [a, *], Yabin Huang [a], Hongxin Dang [a, b]

[a] *Key Laboratory of Special Functional Materials, Henan University, Kaifeng 475001, PR China*
[b] *State Key Laboratory of Solid Lubrication, Lanzhou Institute of Chemical Physics, Chinese Academy of Science, Lanzhou 730000, PR China*

Corresponding author. Tel.: +86 378 2193262; fax: +86 378 2867282.    *E-mail address:* zld@henu.edu.cn (Z. Du).

## Abstract

Spectra of octadecylamine (ODA) Langmuir monolayers and egg phosphatidylcholine (PC)/ODA-mixed monolayers at the air–water interface have been acquired. The organization of the monolayers has been characterized by surface pressure–area isotherms. Application of polarized optical microscopy provides further insight in the domain structures and interactions of the film components. Surface-enhanced Raman scattering (SERS) data indicate that enhancement in Raman spectra can be obtained by strong interaction between headgroups of the surfactants and silver particles in subphase. By mixing ODA with phospholipid molecules and spreading the mixture at the air–water interface, we acquired vibrational information of phospholipid molecules with surfactant-aided SERS effect.

## 1. Introduction

As an important part of nanotechnique, monolayer and Langmuir–Blodgett (LB) techniques have been attracting considerable research interest because of their potential practical applications [1,2]. These techniques are among the powerful means of controlling the molecular orientations and packing at molecular level to construct supramolecular assemblies with special structures and properties. LB technique is also an effective method to study biomineralization and molecular aggregated structures in biological cell membranes, and many promising applications in biological area have been reported [3–5]. In order to study the process of molecule organization and directly acquire more important information of aggregation structure in the Langmuir monolayer in situ, many sensitive surface analytical techniques have been utilized including fluorescence microscopy and Brewster-angle microscopy (BAM) [6,7]. All these tech-niques can provide us some direct information upon revealing the Langmuir monolayer structures at the micrometer level. Ignes-mullol and Schwartz had found a hexatic aggregation of Langmuir monolayer due to shear-inducing molecular precession by using BAM [8]. Observation of the phase separation in mixed lipid monolayer was reported by Bohdana and Stephen [9]. Future fundamental investigations of Langmuir layer depend largely on the invention and application of new analytical techniques to acquire more information at the nanometer level and molecular level.

Surface-enhanced Raman scattering (SERS), a branch in vibrational spectroscopy, has become a powerful tool for structural analysis [10,11]. SERS can provide information on single molecular sensitivity for surface-confined molecules [12]. It is very sensitive to the conformations of individual assembling molecules and their molecular environments, and SERS can also be used to study the structures of monolayer in biologically important aqueous media with negligible spectral interference from water. Recently, several research groups have combined the unique features of SERS with LB technique to study Langmuir monolayer [13–15].

Compared to other SERS-active systems, silver citrate hydrosol is a convenient and simple (both in preparation and manipulations) SERS-active substrate, which is widely used in the study of supramolecular systems in aqueous solutions [16,17]. In present study, we directly use silver colloids as the subphase for the preparation of Langmuir monolayer. The main goal is to obtain SERS spectra of different amphiphilic molecules on the silver colloids using the SERS-active properties of top colloidal silver aggregates. We have obtained SERS spectra of ODA Langmuir monolayer using silver colloids subphase. The microscopic observations prove that the strong interaction of colloidal silver particles with ODA molecules is the main source of the SERS signal. By mixing ODA molecules with egg phosphatidylcholine (PC), we obtained the vibrational information of PC headgroups with the surfactant-aided SERS effect.

## 2. Experimental

### 2.1. Materials

Octadecylamine (ODA) and egg phosphatidylcholine (Tokyo Kasei Chemical Co.) were used as received. ODA and PC were dissolved in chloroform (A.R. grade) at a concentration of $1 \times 10^{-3}$ M to form monolayers on the subphase. Other chemicals were of reagent grade or better and used as received. Water for all experiments was purified to a resistance of 18 MΩ cm.

### 2.2. Ag hydrosol preparation

Colloidal silver nanoparticles were prepared in an aqueous solution by the reduction of silver nitrate with sodium citrate following standard procedure reported by Lee and Meisel [16]. AgNO$_3$ (90 mg) was dissolved in 500 ml of water and heated to boiling; 10 ml of 1% solution of sodium citrate was added dropwise to the boiling solution under vigorous stirring. The solution was kept boiling for 90 min. Then it was cooled and stored at room temperature. Layer of citrate ions absorbed on silver particles leads to an overall negative surface charge and produces colloids that are stable over several weeks at room temperature. The prepared hydrosol showed one absorption peak at 416 nm. The silver colloidal particles obtained have sizes of ca. 50–70 nm in diameter as determined by TEM. Silver colloidal solution was incubated for 1 h with NaCl (ca. 0.2 M final concentration) in order to increase SERS activity [17].

### 2.3. Isotherm measurement

A laboratory-built Langmuir minitrough (200 mm × 65 mm × 7 mm) was used for the preparation of Langmuir films. The trough and barrier were made of PTFE (polytetrafluoroethylene), and surface pressures were measured using a Wilhelmy balance. All movements of balance and barrier were controlled by a computer. The subphase temperature was held constant at $22 \pm 1$ °C during all experiments.

### 2.4. Raman spectral measurement and image observation in situ

Schematic diagram of the laboratory-built set-up for in situ monitoring is shown in Fig. 1. Raman spectra were recorded by using a Jobin Yvon U1000 scanning Raman spectrometer with an Olympus BH2 microscope stage and a cooled photomultiplier (C31034/528 ADS-120) detector. A 514.5 nm exciting line of an Ar$^+$ laser (Spectra-Physics, Model 165, 10 mW at the sample position) irradiated vertically on Langmuir monolayers. Raman scattering from Langmuir layers was collected using conventional 180° backscattering geometry. The monochromator slits were set to 500 μm. Data acquisition was performed using Prism software. Optical microscopy images of the air–water interface were captured with a CCD camera and displayed on computer monitor screen during LB experiments. These experiments were carried out using Olympus BH2 polarized optical microscope within the Raman set-up (observed under a 50× objective for Fig. 6, under 10× objective for Fig. 3).

### 2.5. Procedure

Firstly, LB trough was filled with prepared silver colloids about 90 ml as subphase and allowed to equilibrate for several minutes and then thoroughly eliminated from surface contaminations. Langmuir film was prepared by spreading a chloroform solution of lipid surfactants (27 μl) onto silver colloids subphase with a 50 μl microsyringe. Chloroform was allowed to evaporate for 10 min. Then compression was initiated at a speed controlled by a computer until desired surface pressure was reached. In the case of mixed monolayer, solid ODA and PC mixture (ODA:PC = 1:1 mol/mol) was dissolved in chloroform at a concentration of $1 \times 10^{-3}$ M for preparation of monolayer. After relaxing of Langmuir film for 10 min, Ar$^+$ 514.5 nm exiting lines irradiated perpendicularly on the air–water interface.

## 3. Results and discussion

### 3.1. ODA Langmuir monolayer

#### 3.1.1. π–A isotherms

ODA was chosen as initial lipid surfactant for our study due to its simple structure and regular use. Fig. 2 shows π–A isotherms of ODA on pure water and silver colloids. Compared with the isotherm on pure water, a shift of ca. 10 Å$^2$ per molecule to larger values is observed for isotherm on the silver colloids. The area per molecule value extrapolated to $\pi = 30$ mN/m for ODA Langmuir monolayer on silver colloids is ca. 24 Å$^2$. As we know, layer of citrate ions absorbed on silver particles leads to a negative overall surface charge for silver particles in subphase and ODA molecules show a positive charge in the air–water interface [18]. Some papers have reported strong interactions between headgroups of ODA and colloidal silver particles [19,20]. The increase of the molecule area of ODA monolayer indicates that there is a strong electrostatic interaction between the protonated amine groups in Langmuir monolayer and negatively charged silver nanoparticles in subphase. Fig. 2b shows a scheme of ODA monolayer on silver colloid subphase.

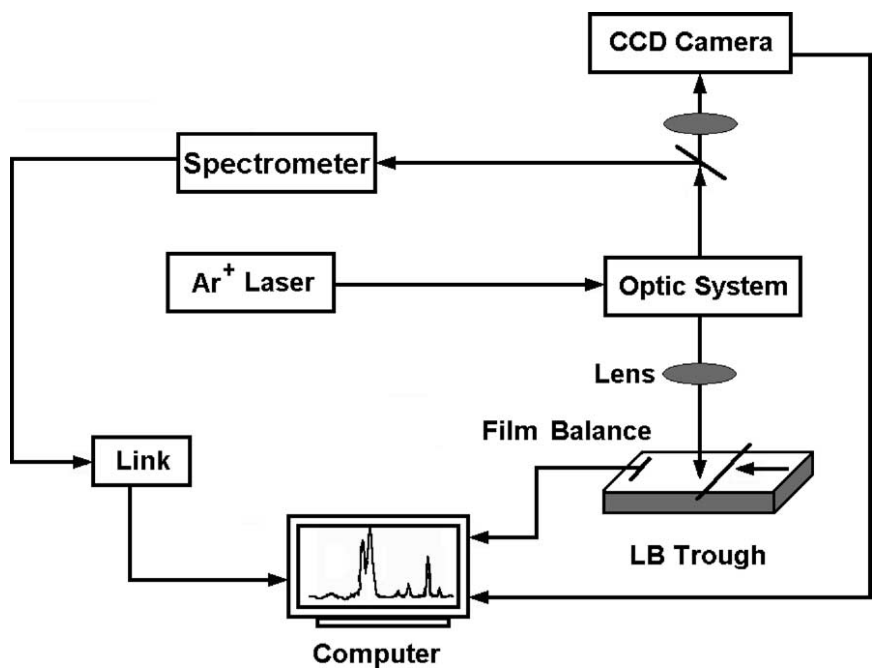

Fig. 1. Experimental set-up for obtaining Raman spectra of Langmuir monolayers and process monitoring.

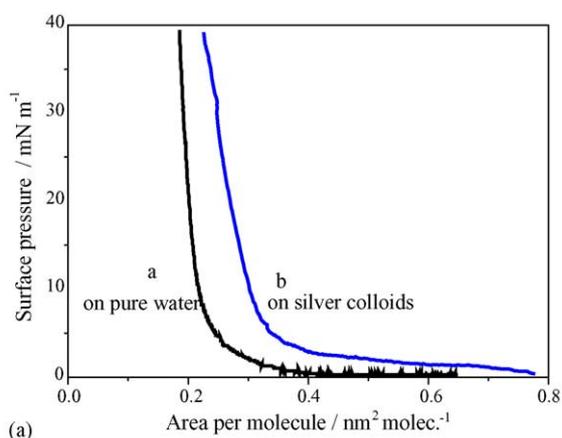

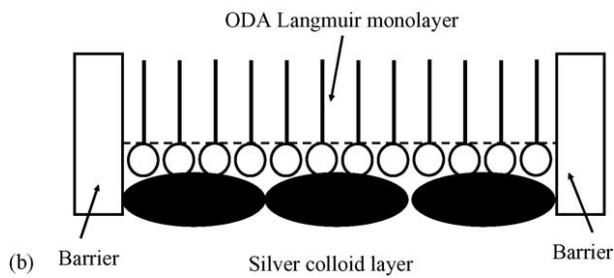

Fig. 2. Surface pressure–area isotherms (a) and scheme (b) of ODA on pure water and on silver colloids at 22 °C.

### 3.1.2. Observation of images in situ

Fig. 3a presents a direct microscopic observation of the surface of silver colloids before spreading of any surfactant. Generally, a smooth surface can be seen for subphase. After spreading of chloroform solution of ODA at the air–water interface and compressing of ODA film to the pressure of 30 mN/m, we obtain the image of subphase surface as shown in Fig. 3b. With close observation of Fig. 3b, it can be seen that a closely packed colloidal silver layer is formed. In process of formation for ODA Langmuir monolayer, lots of colloidal silver particles were absorbed below the headgroups of ODA molecules. Coalescence of silver particles would occur when the growing particles attained a sufficient size and linked laterally under ODA monolayer. Finally, a silver colloid layer formed under ODA Langmuir monolayer.

### 3.1.3. SERS of ODA Langmuir monolayer

In our Raman spectral measurements, we cannot detect any normal Raman vibration information of ODA Langmuir monolayer on pure water because of the small number of molecules and the weakness of the signal. In case of ODA Langmuir monolayer on silver colloids subphase, SERS spectra of the ODA Langmuir monolayer can be readily observed with the layer of silver colloids formed under ODA film as shown in Fig. 4. The band frequencies between 200 and 1800 cm$^{-1}$ of ODA Langmuir monolayer (spectrum b) are in good agreement with those shown for solid ODA (spectrum a), although the relative intensities are different because of selection rules associated with surface enhancement on metal substrates. The bands at 230 cm$^{-1}$ in spectrum b are assigned to Ag–N stretching mode [21,22], which is characteristic frequency involving vibration of chemical groups bond to Ag particles or aggregates. It indicates that there is a chemical adsorptive bond between silver particle and ODA molecule via NH$_2$ group. In conclusion, the presence of SERS of Langmuir monolayer mainly attributed to strong interaction between headgroups of ODA and silver particles.

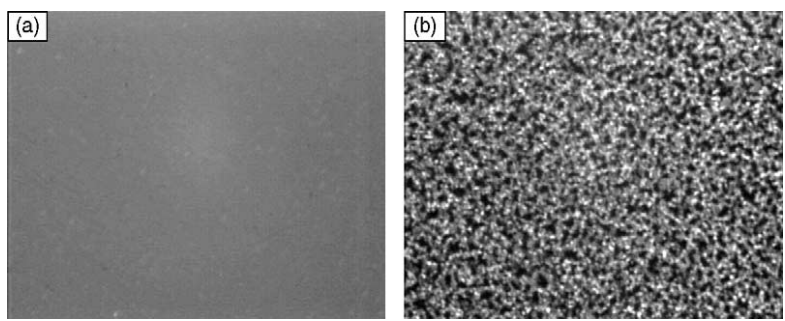

Fig. 3. Microscopic observations of (a) silver colloids surface and (b) ODA Langmuir monolayer at surface pressure of 30 mN/m on silver colloids. Images are 150 μm × 110 μm in size.

For ODA Langmuir monolayer, the bands in the region of 800 and 1600 cm$^{-1}$ are relatively similar to those present in spectra of solid ODA with little frequency shift and changes of intensity. Vibrational assignments are based on those for other long alkyl chain systems [23]. The presence of *trans* $v_a(CC)_T$ and $v_s(CC)_T$ modes at ca. 1060 and 1130 cm$^{-1}$, methylene twisting mode, (t(CH$_2$)) at 1294 cm$^{-1}$ and methyl rocking mode at ca. 888 cm$^{-1}$ suggest that alkyl chain portion of the ODA Langmuir monolayer is ordered and carbon backbone of ODA is largely in an all-*trans* conformation

[23–25]. This conclusion is consistent with previous SERS works of alkyl chains SAM monolayers on Ag particles and Ag-polished substrate [24,25]. The bands at ca. 1432 and 1452 cm$^{-1}$ are assigned to CH$_2$ scissoring modes and bands of 1390 cm$^{-1}$ corresponds to CH$_2$ wagging mode (w(CH$_2$)) [13,26–28]. The broad bands at 1590 cm$^{-1}$ are assigned to graphitic carbon originating from a photochemical reaction between CO$_2$ present in air or water and the silver layer [29,30].

When ODA molecules were compressed, NH$_2$ groups of ODA may be stuck on the silver particles layer. Comprehensive interactions may exist in compressed monolayer, which mainly control the structure of Langmuir monolayer and influence the SERS of Langmuir monolayer. Based on the selection rules of SERS for molecules adsorbed on metal surfaces, molecular vibrations involving motions perpendicular to the surface should be enhanced in SERS spectra, while those involving motions parallel to the surface weakened. Therefore, SERS can be used for qualitative determination of molecular orientation at surfaces [31]. The changes of relative intensity for some bands in Fig. 4b show significant orientational effects on SERS of ODA Langmuir monolayer. In particular, drastic increase in the intensities of $v_s(CC)_T$ at 1130 cm$^{-1}$ and w(CH$_2$) mode of 1390 cm$^{-1}$ and decrease of t(CH$_2$) mode at 1294 cm$^{-1}$ in SERS spectra indicate that alkyl chains of ODA molecules are perpendicular to the surface of Ag films [32], which is consistent with microscopic observation of ODA Langmuir monolayer.

Fig. 5 shows Raman spectra in the C–H stretching region of solid ODA and ODA Langmuir monolayer. We can note that the band shapes and frequency positions are mainly the same in both spectra. The bands at 2846, 2881 and 2900 cm$^{-1}$ in spectrum of Langmuir monolayer are similar to those present on the solid spectrum and can be assigned, respectively, to the symmetric and asymmetric CH$_2$ stretching modes [33,34]. It can be seen that there is a relative decrease of the $v_a(CH_2)$ intensity with respect to the $v_s(CH_2)$ in spectrum b. The relative intensities of the $v_a(CH_2)$ and $v_s(CH_2)$ can be associated with a conformational and structural ordering of the monolayer, which indicates that ODA Langmuir monolayer is in a less crystalline state than in solid phase [13,24]. The ODA Langmuir film is partially ordered and in a liquid-crystalline state because of strong interaction between headgroups of surfactants and silver nanoparticles. This is accordant with the results concluded from above $\pi$–A isotherms.

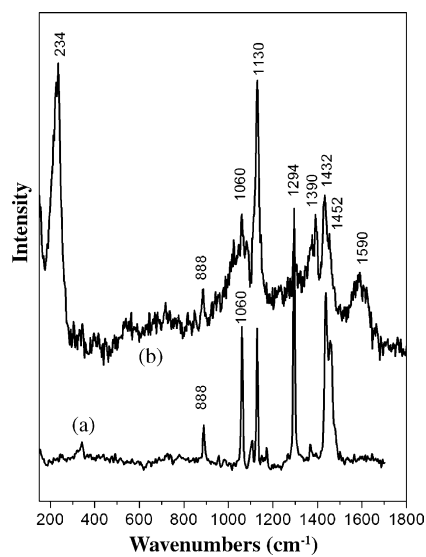

Fig. 4. Raman spectra of (a) solid ODA and (b) ODA Langmuir monolayer at surface pressures of 30 mN/m on silver colloids between 200 and 1800 cm$^{-1}$.

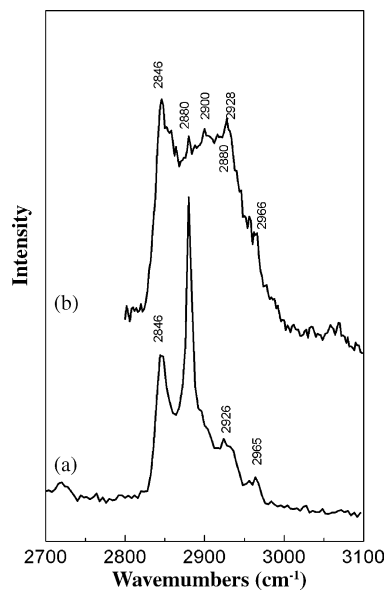

Fig. 5. Raman spectra in the $v$(C–H) region for (a) solid ODA and (b) ODA Langmuir monolayer at surface pressure of 30 mN/m on silver colloids.

### 3.2. PC monolayer and PC/ODA-mixed Langmuir monolayer

#### 3.2.1. Image and $\pi$–A curve of monolayers

PC monolayer has long been regarded as a desirable model of biomembrane and has potential applications in medical area. To appropriately use phospholipid as a model of biomembranes and a carrier in drug delivery, systematic knowledge of the structure and aggregation processes is necessary. In order to study the properties of phospholipid Langmuir layer, monolayer of PC has been deposited on silver colloid subphase by LB technique. Fig. 6a shows direct microscopic observation of subphase surface after Langmuir monolayer was formed. From the image, we can see that surface of silver colloids is smooth without any formation of silver aggregates underneath PC monolayer, which is very similar to the image obtained form surface of silver colloid subphase without depositing any surfactant (Fig. 3a). The headgroup of PC molecule is electrically neutral on silver colloids subphase in our experimental conditions. The microscopic observations indicate that headgroups of PC molecules are weakly adsorbed on colloidal silver particles or interaction between PC molecules and silver colloids in subphase does not exist. In isotherms experiments, we have also observed no change in isotherm of PC monolayer on silver colloids subphase and pure water subphase.

According to the knowledge obtained from experiments mentioned above, we mixed ODA and PC (ODA:PC = 1:1 mol/mol) in Langmuir film in order to obtain vibrational information of PC film with the surfactant-aided SERS effect [35]. Fig. 6b shows direct microscopic observation of ODA/PC-mixed monolayer, many colloidal silver aggregates appear at the air–water interface. The image indicates that ODA molecules in the mixed film interacted with silver nanoparticles and formed colloidal silver aggregates. Fig. 7b shows a scheme of ODA/PC-mixed monolayer on silver colloid subphase. Apparently, the formation of colloidal silver aggregates at subphase surface is mainly caused by interaction of ODA molecules and silver nanoparticles.

Surface pressure–area isotherm for PC/ODA-mixed film on silver colloids is shown in Fig. 7a. The isotherm shows that PC/ODA-mixed molecules can form a preferable monolayer on silver colloids. Compared with isotherm obtained for ODA on silver colloid, isotherm of mixed molecules shows a large tilt angle. It shows PC/ODA-mixed monolayer is more compressible than ODA monolayer. This can be explained by the presence of PC molecules which having good properties of liquid crystalline. The average molecular area per molecule value extrapolated to $\pi = 30$ mN/m for mixed Langmuir monolayer is ca. 25 Å$^2$. By extrapolating linear part of the curve to zero surface pressure, a limiting molecular area of 41 Å$^2$ is obtained. We know that a pure PC molecule has two alkyl chains and a limiting area of 56 Å$^2$. If ODA and PC formed a single monolayer, by calculation we obtain that the average molecular area and limiting molecular area should be 30 and 44 Å$^2$, respectively (here, a limiting area of 32 Å$^2$ for ODA is used which can be obtained from Fig. 2.).

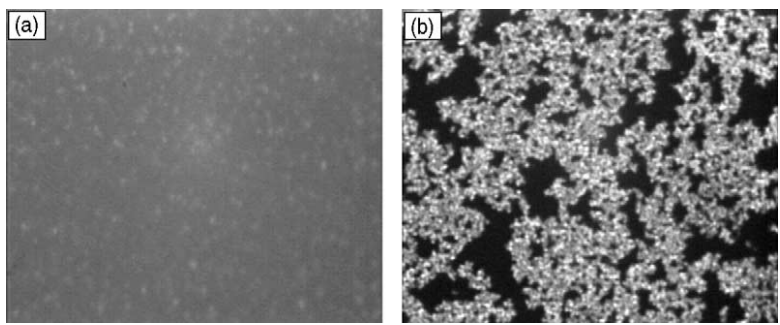

Fig. 6. Microscopic observations of (a) PC Langmuir layer and (b) ODA/PC-mixed Langmuir monolayer at surface pressure of 30 mN/m on silver colloids. Images are 150 μm × 110 μm in size.

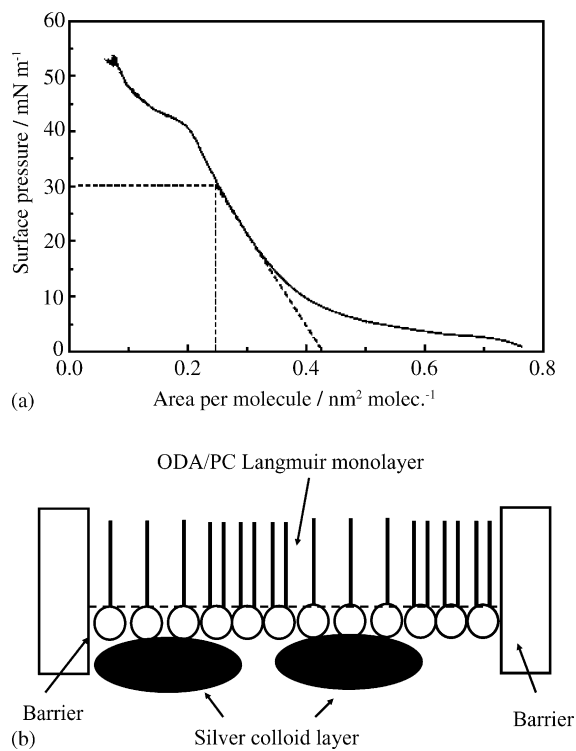

ODA/PC Langmuir monolayer

Barrier          Silver colloid layer          Barrier
(b)

Fig. 7. Surface pressure–area isotherm (a) and scheme (b) of ODA/PC-mixed Langmuir monolayer on silver colloids.

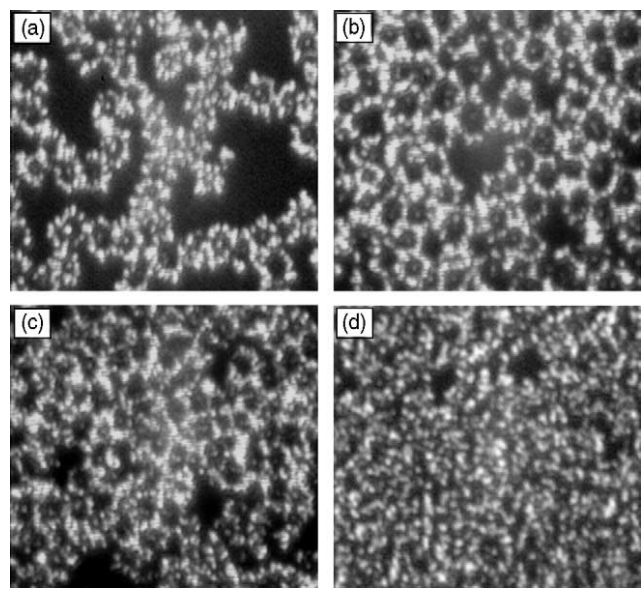

Fig. 8. Microscopic observations of ODA/PC-mixed Langmuir monolayer obtained during continuous compression at the surface pressures indicated (a) 10, (b) 20, (c) 30 and (d) 40 mN/m. Images are 30 μm × 24 μm in size.

This reduction of molecular area in experiments than that in calculation suggests that tighter packing is occurred in the ODA/PC-mixed monolayer.

As can be seen from Fig. 2a and Fig. 7a, ODA Langmuir monolayer and ODA/PC monolayer have a similar average molecular area of ca. 25 Å² under the pressure of 30 mN/m. Because an alkyl chain occupies ca. 18 Å² and ODA has only one alkyl chain per molecule, while ODA/PC has three alkyl chains per unit, we considered that alkyl chains are not closely packed in pure ODA Langmuir monolayer and there is a much closer packing in ODA/PC-mixed film than in pure ODA film. This difference in structure and packing density has a significant influence for the acquisition of Raman spectra in the following step.

### 3.2.2. Aggregation of ODA/PC-mixed monolayers

With CCD imaging system, we can monitor the whole compression process of Langmuir layers directly. Fig. 8 shows images of ODA/PC Langmuir layer on silver colloids in the continuous compression process acquired at pressures of 10, 20, 30 and 40 mN/m, respectively. From above discussion of Fig. 3, we can see that bright spots in the images present formation of colloidal silver aggregates underneath ODA molecules. We consider that bright network structures indicated aggregation areas of ODA molecules and dark areas in the image are occupied by PC aggregation.

From the π–A curve of ODA/PC (see Fig. 7) at a pressure of 10 mN/m, the film is in liquid-expanded state, but ODA film has induced aggregation of Ag particles as shown in Fig. 8a; at a pressure of 20 mN/m, the film is in liquid-condensed phase, it can be seen that a circle-like network structure formed at the air–water interface due to different aggregation behavior of PC and ODA molecules (shown as Fig. 8b). With pressure increasing, previous network structure was continuously compressed and most of circle-like network structure was distorted at a pressure of 30 mN/m (Fig. 8c). In Fig. 8d, the Langmuir film appears in solid phase with further increase of pressure. We can still see aggregation of PC and ODA dispersed in different phase. By this way, we can directly obtain in situ the information of "domain" of mixed monolayer. The detailed study will be reported in another paper.

### 3.2.3. SERS of ODA/PC-mixed monolayers

In Fig. 9, SERS spectra of mixed ODA/PC (ODA:PC = 1:1 mol/mol) Langmuir layer (spectrum a) is comprared with SERS spectra of ODA Langmuir monolayer (spectrum b). In the low frequency region of spectrum a, the band at 230 cm⁻¹ can be assigned to Ag–N stretching vibration of Ag–ODA complex, and intensity is weaker than that in Fig. 9b because of the amount decrease of ODA. Characteristic bands of alkyl chains appear in the frequency region of 800–1600 cm⁻¹, which include $v$(C–C) stretching vibrations in the 1000–1130 cm⁻¹ and C–H bending modes in 1200–1500 cm⁻¹. For ODA/PC-mixed Langmuir monolayer, the bands at 1000 cm⁻¹ correspond to symmetric $PO_3{}^{2-}$ stretching mode [36]. The bands at 1084 and 1236 cm⁻¹ are assigned to symmetrical and asymetrical $PO_2{}^-$ stretching modes of phosphate groups in PC molecules [37,38]. From Fig. 9c, we cannot find any signal of PC monolayer in the spectrum. However, by mixing PC with ODA, we acquired the vibrational information of PC headgroups, which can be attributed to strong interaction between ODA molecules and silver nanoparticles. The above assignments are reasonable because in this case the ODA and PC molecules have a tighter packing structure, which can directly observe from surface pressure–area isotherm and microscope images of mixed monolayers. Moreover the ensuing alkyl chain aggregates would exhibit increased interchain ordering characteristics. As shown, the w(CH$_2$) band of 1390 cm$^{-1}$ is significantly strong in ODA/PC-mixed monolayer. This enhancement of vibrational coupling mainly originates from intermolecular chain–chain interactions.

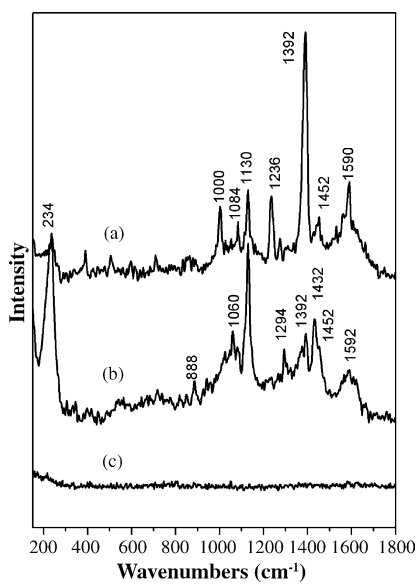

Fig. 9. Raman spectra of (a) ODA/PC Langmuir monolayer, (b) ODA Langmuir monolayer and (c) PC monolayer at surface pressures of 30 mN/m on silver colloids at 200–1800 cm$^{-1}$.

## 4. Conclusions

In present paper, we reported acquirement of SERS spectra from Langmuir monolayers of ODA using a silver colloids subphase. The microscopic observations provides direct evidence for the formation of particulate silver layer prove that silver aggregates under ODA Langmuir film induced by headgroups of ODA molecules are main source of the SERS signal acquisition. SERS data indicate that SERS enhancement of ODA Langmuir monolayers at the air–water interface results from strong interaction between headgroups of ODA molecules and silver nanoparticles in subphase. Spectra for ODA Langmuir monolayers suggest that complex interactions exist in the monolayers, including hydrophobic interaction in fatty chains of ODA and interaction between headgroups of surfactants and metal nanoparticles in subphase.

Pure PC Langmuir monolayers on silver colloids do not show any SERS signal mainly because interaction between PC molecules and the silver colloid surface is weak. By mixing ODA with PC molecules, from microscopic observations on compression process, we can see phase separation behavior in the ODA and PC mixed film on silver colloids. SERS spectra of PC molecules can be obtained from PC/ODA mixed lipid monolayers by surfactant-aided SERS effect. This surfactant-aided methodology has been extended to obtain SERS spectra of other surfactants. Work based on such systems could expand SERS application in the study of Langmuir monolayers at the air–water interface.

## Acknowledgement

This work was supported by Natural Science Foundation of China (Nos. 90306010 and 20371015) and State Key Basic Research "973" Plan of China (No. 2002CCC02700).